\documentclass[twocolumn,showpacs,preprintnumbers,amsmath,amssymb,floatfix]{revtex4}
\usepackage{graphicx}% Include figure files
\usepackage{afterpage}% help with floats
\usepackage{amsmath}% for help with equations
\usepackage{dcolumn}% Align table columns on decimal point
\usepackage{bm}% bold math

\begin{document}

\preprint{APS/123-QED}

\title{Detailed studies of non-linear magneto-optical resonances at $D_1$ excitation of $^{85}$Rb and $^{87}$Rb for partially resolved hyperfine $F$-levels} 

\author{M.~Auzinsh}
\email{Marcis.Auzins@lu.lv}
\author{R.~Ferber}%
\author{F.~Gahbauer}%
\author{A.~Jarmola}%
\author{L.~Kalvans}%
\affiliation{
The University of Latvia, Laser Centre, Rainis Blvd., LV-1586 Riga, Latvia}%

%\affiliation{

\date{\today}% It is always \today, today,

\begin{abstract}
Experimental signals of non-linear magneto-optical resonances at $D_1$ excitation of natural rubidium in a vapor cell have been obtained and described with experimental accuracy by a detailed theoretical model based on the optical Bloch equations. The $D_1$ transition of rubidium is a challenging system to analyze theoretically because it contains transitions that are only partially resolved under Doppler broadening. The theoretical model took into account all nearby transitions, the coherence properties of the exciting laser radiation, and the mixing of magnetic sublevels in an external magnetic field and also included averaging over the Doppler profile. Great care was taken to obtain accurate experimental signals and avoid systematic errors. The experimental signals were reproduced very well at each hyperfine transition and over a wide range of laser power densities, beam diameters, and laser detunings from the exact transition frequency.
The bright resonance expected at the $F_g=1\rightarrow F_e=2$ transition of $^{87}$Rb has been observed. A bright resonance was observed at the  $F_g=2\rightarrow F_e=3$ transition of $^{85}$Rb, but displaced from the exact position of the transition due to the influence of the nearby  $F_g=2\rightarrow F_e=2$ transition, which is a dark resonance whose contrast is almost two orders of magnitude larger than the contrast of the bright resonance at the $F_g=2\rightarrow F_e=3$ transition. Even in this very delicate situation, the theoretical model described in detail the experimental signals at different laser detunings. 

\end{abstract}
%32.60.+i Zeeman and Stark effects
%32.80.Xx Level crossing and optical pumping
%32.10.Fn Fine and hyperfine structure
\pacs{32.60.+i,32.80.Xx,32.10.Fn}% PACS, the Physics and Astronomy
\maketitle

\section{\label{Intro:level1}Introduction}
Dark~\cite{Alzetta:1976,Schmieder:1970} and bright~\cite{Dancheva:2000} non-linear magneto-optical resonances associated with atomic state coherences~\cite{Arimondo:2007} and strongly related to the ground state Hanle effect~\cite{Lehmann:1964} (see~\cite{Strumia} for a review) are a particular manifestation of a broad class of non-linear magneto-optical phenomena (see, for example, \cite{Budker:2002,Alexandrov:2005} for a review). Dark resonances in a scan of laser induced fluorescence (LIF) versus magnetic field  arise when at zero magnetic field coherences are induced among ground state magnetic sublevels and a quantum superposition state is created that is not coupled to the exciting laser field. Bright resonances arise when atoms at zero magnetic field are prepared in a superposition of ground state sublevels that is strongly coupled to the excited state. These resonances have a variety of applications, such as in optical magnetometers~\cite{Budker:2007} and optical switches~\cite{Yeh:1982}, and they have been studied extensively in many systems both theoretically and experimentally (see, for example,~\cite{Lehmann:1964, Picque:1978, Renzoni:1997, Ling:1996, Dancheva:2000, Renzoni:2001a, Renzoni:2001b, Alzetta:2001, Andreeva:2002, Papoyan:2002, Alzetta:2004, Gateva:2005, Andreeva:2007}). However, accurately modelling the profiles of bright and dark resonances in systems that are unresolved under Doppler broadening remains a challenging enterprise. 
There are two ways to address the issue of Doppler broadening in a detailed study of these resonances in alkali vapors: (1) compensating for the lack of spectral resolution by using a detailed theoretical model or (2) having recourse to an extremely thin cell whose width is comparable to the wavelength of the laser radiation, and which has sub-Doppler resolution~\cite{Andreeva:2007}.
We chose the first approach and aimed to demonstrate that it is possible to achieve with experimental accuracy a quantitative description of the shape of all bright and dark resonances of the $D_1$ transition of $^{85}$Rb and $^{87}$Rb by means of a straightforward theoretical model based on the optical Bloch equations with minimal recourse to adjustable parameters.
In particular, we wanted to study the dependence of the resonance shapes, in particular the resonance contrast and width, on the laser power density, laser frequency, and laser beam diameter. The laser beam diameter is related to the transit relaxation time, which is critical for determining the resonance profile. Previous works did not study how the resonance shape varies with beam diameter. 

The ground state Hanle effect was first observed by Lehmann and Cohen-Tannoudji in 1964~\cite{Lehmann:1964}. In 1978, Piqu\'e presented measurements of dark resonances in the $F_g=2\rightarrow F_e=1$ transition of the $D_1$ line in a beam of sodium atoms with negligible Doppler broadening, which agreed quite well with calculations based on the optical Bloch equations~\cite{Picque:1978}. Renzoni et al.~\cite{Renzoni:1997} studied experimentally and theoretically all transitions of the $D_1$ sodium line in a beam of sodium atoms, but saw only dark resonances. Bright resonances were first observed in 2000 in the $D_1$ and $D_2$ transitions of rubidium by Dancheva et al.~\cite{Dancheva:2000} in a vapor of rubidium atoms, but the study did not include a model-based analysis of the results. Papoyan et al.~\cite{Papoyan:2002} studied bright and dark resonances in the $D_2$ transition of cesium, but their theoretical model based on the optical Bloch equations predicted a bright resonance for the $F_g=4\rightarrow F_e=3,4,5$ transition, which appeared to be dark in the experiment.  Alzetta et al.~\cite{Alzetta:2004} excited the $D_1$ transition of sodium in a vapor cell with a broadband multi-mode laser to demonstrate dark resonances with 100\% contrast and modelled these resonances with rate equations. However, the model was not used to predict the shape of the resonances. Gateva et al.~\cite{Gateva:2005} measured the widths of the dark resonances at the $F_g=2\rightarrow F_e=1$ transition of the $D_1$ line of $^{87}$Rb and compared the widths to predictions based on an analytical expression obtained for a Doppler broadened three-level system in the linear approximation. Most recently~\cite{Andreeva:2007}, experiments were performed on the $D_2$ transition of cesium in an extremely thin cell, but some resonances that should have been bright appeared dark for as yet unknown reasons, and the resonances are rather broad. The most advanced previous attempt to apply a detailed model magneto-optical resonances in alkali vapors was made by Andreeva et al.~\cite{Andreeva:2002}, who worked on the cesium $D_2$ transition, which is unresolved under Doppler broadening. Their model took into account all nearby hyperfine transitions and averaged over the Doppler profile and over the distribution of transit times through the laser beam. Nevertheless, the theoretical and experimental contrast of these resonances as a function of laser power density presented in this study differed by about a factor of two. Moreover, the paper presented a comparison of theoretical and measured resonance contrast as a function of laser power density only for one bright resonance and did not study the dependence on beam diameter. One weak point in the model was that it did not take into account the coherence properties of the laser radiation in sufficient detail. 

\begin{figure}[htbp]
	\centering
		\resizebox{\columnwidth}{!}{\includegraphics{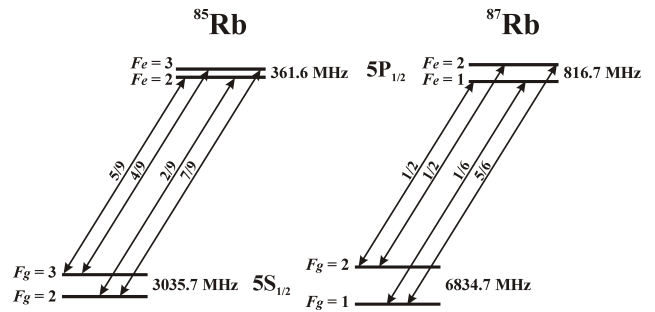}}
	\caption{\label{fig:levels} Hyperfine level structure and transitions of the  $D_1$ line of rubidium. The fractions on the arrows indicate the relative transition strengths~\cite{steck:rubidium85,steck:rubidium87}.}
\end{figure}

\begin{figure}[htbp]
	\centering
		\resizebox{\columnwidth}{!}{\includegraphics{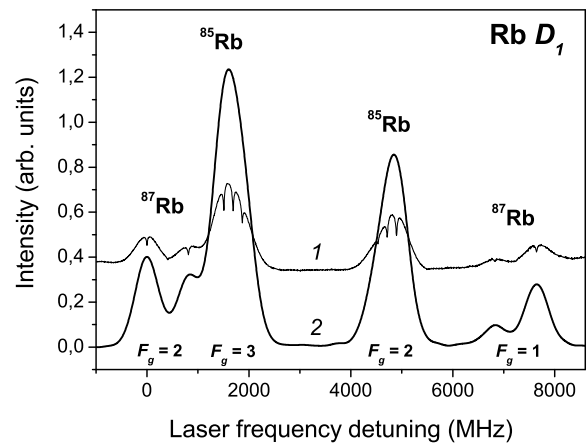}}
	\caption{\label{fig:LIF_excitation} Spectrum of Rb $D_1$ line. Curve 1---saturated absorption spectrum; curve 2---LIF excitation spectrum at 10 mW/cm$^2$ excitation.}
\end{figure}

In this paper we endeavored to show that the optical Bloch equations can model correctly bright and dark resonances in vapor cells when the model takes into account nearby hyperfine levels and magnetic sublevel mixing, averages over the Doppler profile, and carefully includes the coherence properties of the laser radiation. We first had applied such a detailed model to the $D_1$ line of cesium in an atomic vapor~\cite{Auzinsh:2008}. Cesium $D_1$ excitation has the simplifying advantage that the energy splitting between the excited state hyperfine levels exceeds the Doppler broadening. 
After having gained confidence in the model's suitability for describing a simple situation, we now report on studies of a more complex system, which tested more rigorously the applicability of the model under a wide variety of experimental conditions. We chose the $D_1$ line of natural rubidium, whose excited state hyperfine levels can be resolved only partially under Doppler broadening. Figure~\ref{fig:levels} shows the level scheme of the $D_1$ line in $^{85}$Rb and $^{87}$Rb. The energy difference between the excited state hyperfine levels is 361.6 MHz for $^{85}$Rb and 816.7 Mhz for $^{87}$Rb. The full width at half maximum (FWHM) of the Doppler broadening for rubidium at room temperature is about 500 MHz. The arrows represent the possible transitions between ground and excited states, and the fractions next to the arrows represent the relative transition strengths. Figure~\ref{fig:LIF_excitation} shows a LIF excitation spectrum of the $D_1$ line of natural rubidium obtained with the same experimental equipment as the other results described in this paper. The exact positions of the resonances can be seen from the simultaneously recorded saturated absorption spectrum. One can see that the two excited state hyperfine levels of  $^{87}$Rb are almost resolved under Doppler broadening, while the hyperfine levels of  $^{85}$Rb are rather poorly resolved, which is consistent with the energy level separations given in Figure~\ref{fig:levels} and the Doppler width of rubidium at room temperature. This system is interesting, because it is expected to contain six dark and two bright resonances, of which the bright resonance at the $F_g=1\rightarrow F_e=2$ of $^{87}$Rb has never before been observed as far as we know.

\section{\label{Experiment:level1}Experimental Description}
\begin{figure}[htbp]
	\centering
		\resizebox{0.5\columnwidth}{!}{\includegraphics{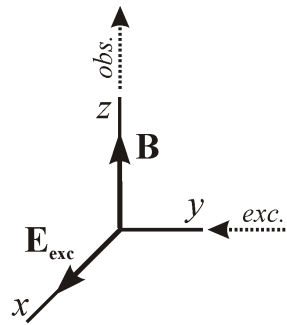}}
	\caption{\label{fig:geometry} Experimental geometry. The relative orientation of the laser beam (\textit{exc}), laser light polarization (\textbf{E$_{exc}$}), magnetic field (\textbf{B}), and observation direction (\textit{obs}) are shown. 
}
\end{figure}

	The experimental geometry is shown in Figure~\ref{fig:geometry}, which depicts the relative orientation of the laser beam (\textit{exc}), the laser radiation's linear polarization vector (\textbf{E$_{exc}$}), the magnetic field (\textbf{B}), and the observation direction (\textit{obs}). The experimental setup and auxiliary equipment are similar to what was used in~\cite{Auzinsh:2008}. In the present experiment, natural rubidium was confined at room temperature in cylindrical glass cells. The laser was a DL 100 external cavity single mode diode laser with a wavelength of 794.3 nm and a typical line width of a few Megahertz, which was produced by Toptica, A.~G., of Graefelfing, Germany. Two different cells were used in the experiments. One was produced by Toptica, A.~G., and made of Pyrex with optical quality windows with a diameter of 25 mm.  The other cell was made in our laboratory and had a diameter of 35 mm. The cell was placed at the center of a three-axis Helmholtz coil system. Two pairs of coils compensated the laboratory magnetic field, while the third pair of coils was used to scan the magnetic field in the observation direction. The current in this third pair of coils was scanned with a Kepco BOP-50-8M bipolar power supply that was controlled by an analog signal from a computer. For some of the experiments, part of the laser radiation was diverted to the second cell, which was used to obtain a saturated absorption spectrum, from which the exact position of the transitions were determined. In addition to the National Instruments 6024E data acquisition card, an Agilent DSO5014A oscilloscope was used in some of the experiments for data acquisition. Both data acquisition systems produced similar results. 

	The laser induced fluorescence was detected by means of a Thorlabs FDS-100 photodiode as a function of magnetic field for all hyperfine transitions of the two isotopes of rubidium (see Fig.~\ref{fig:levels}). No polarizers were used in the LIF observation. Signals were obtained for different laser power densities (between 5 mW/cm$^2$ and 1000 mW/cm$^2$) and laser beam diameters (between 0.09 mm and 2.3 mm). The beam diameter was assumed to be the full width at half maximum (FWHM) of the beam profile, which was determined by means of a Thorlabs BP104-VIS beam profiler. The beam profile was roughly a symmetrical Gaussian. 

Signals were obtained by employing a double scanning technique~\cite{Andreeva:2002} in which the laser frequency was scanned slowly across a transition or pair of adjacent transitions while the magnetic field was scanned more rapidly following a saw-tooth shape from negative to positive values. The laser frequency changed by about 10--20 MHz per second. The typical laser frequency scan lasted about 2--3 minutes, whereas the typical period for the magnetic field scan was 1 second. In this manner, a series of resonance signals at laser frequencies differing by 10--20 MHz could be obtained. For some observations the laser's frequency was maintained at the maximum of the fluorescence signal at a given transition while the magnetic field was  scanned several times, and the scans were averaged. The laser was not actively stabilized, but its frequency was monitored with a High Finesse WS-7 wavemeter and was found to vary by a quantity on the order of 10 MHz during the time of a typical measurement. 

	In addition to the signal, each measurement included a certain amount of background. Background from scattered laser light could be determined easily by tuning the laser off resonance. However, the background associated with scattered LIF  could not be readily identified. It was assumed that this background accounted for a fixed percentage of the signal for a given vapor cell. The experimental signals from the home-made cell matched the calculated curves when the signals from the homemade cell were assumed to contain about 20\% background from scattered LIF. Because of the improved transparency of the high optical quality glass in the cell from Toptica, A.~G., the component of the background from scattered LIF was so small that the theoretical and experimental curves agreed without assuming any contamination from scattered LIF. With this cell, it was no longer necessary to consider the background from scattered LIF to be an adjustable parameter in the theoretical treatment of the signals. Almost all the signals were remeasured with the cell from Toptica, A.G., in order to verify our results. 
%These percentages, one for each cell, were determined by fitting experimental results with theoretical calculations for a data set that consisted of many resonances that had been recorded for different transitions, laser power densities, and beam diameters. In the Toptica cell, good agreement with calculation was found even neglecting the background from scattered LIF, whereas in the other cell the background from scattered LIF was estimated to be about 20\% of the signal. 

\section{\label{Theory:level1}Theoretical Model}
The theoretical model is basically the same as the one described in~\cite{Auzinsh:2008}. The model assumes that the atoms move classically and are excited at the internal transitions. Thus the internal atomic dynamics can be described by semi-classical atomic density matrix $\rho$, which also depends parametrically on the classical coordinates of the atomic center of mass. The time evolution of the density matrix $\rho$ is described by the optical Bloch equations (OBE)~\cite{Stenholm:2005}
\begin{equation}
        i\hbar \frac{\partial \rho}{\partial t} = \left[\hat{H},\rho \right] + i \hbar\hat{R}\rho,
        \label{obe}
\end{equation}
where the relaxation operator $\hat{R}$ includes the spontaneous emission rate, which equals the natural transition linewidth $\Gamma$ and the transit relaxation rate $\gamma$ (as the atom-laser interaction takes place in a finite spatial region). The Hamiltonian $\hat{H}$ can be written as follows $\hat{H} = \hat{H}_0 + \hat{H}_B + \hat{V}$, where $\hat{H}_0$ is the unperturbed atomic Hamiltonian, which depends on the internal atomic coordinates, $\hat{H}_B$ is the Hamiltonian of the atomic interaction with the magnetic field, and $\hat{V} = -\hat{\textbf{d}} \cdot \textbf{E}(t)$ is the dipole interaction operator, which includes the electric dipole operator $\hat{\textbf{d}}$ and the electric field of the excitation light $\textbf{E}(t)$. The electric field is described classically with some polarization in the direction of the unit vector \textbf{e}, an electric field amplitude $\varepsilon_{\overline{\omega}}$, a central frequency $\overline{\omega}$, and a fluctuating phase $\Phi(t)$, which causes a finite bandwidth $\Delta\omega$ (FWHM) for a Lorentzian shaped spectral profile:
\begin{equation}
        \textbf{E}(t) = \varepsilon(t)\textbf{e} + \varepsilon^{*}(t)\textbf{e}^{*}
        \label{Efield},
\end{equation} 
\begin{equation}
        \varepsilon(t) = \vert\varepsilon_{\overline{\omega}}\vert e^{-i\Phi(t)-i\left(\overline{\omega}-\textbf{k}_{\overline{\omega}}\textbf{v}\right)t}
        \label{epsilon}.
\end{equation}
The atoms move with definite velocity \textbf{v}, which causes the shift $\textbf{k}_{\overline{\omega}}\textbf{v}$ in
the laser frequency due to the Doppler effect, with $\textbf{k}_{\overline{\omega}}$ being the wave vector of the excitation light.

Then we apply the rotating wave approximation~\cite{Allen:1975} to the OBEs. It results in stochastic differential equations, which can be further simplified by using the decorrelation approach~\cite{Kampen:1976}. The
stochasticity is caused by the random fluctuations of the laser radiation and leads to a finite spectral width of the radiation. In the experiment we observed the light intensity, which is a quantity that is averaged over time intervals that are large in comparison with the characteristic time of the phase-fluctuations. Therefore we perform a statistical averaging of the stochastic differential equations. This is done by taking a formal statistical average over the fluctuating phases and then using the decorrelation approximation~\cite{Blushs:2004}, which is valid for either the random phase jump or the continuous random phase diffusion model. As a result, we eliminate the density matrix elements that correspond to optical coherences and arrive at equations for the Zeeman coherences,
\begin{align}
\dfrac{\partial\rho_{g_ig_j}}{\partial t} =&\bigl(\Gamma_{p,g_ie_m} + \Gamma_{p,e_kg_j}^{\ast}\bigr)\underset{e_k, e_m
}{ \sum }\bigl(d_1^{g_ie_k}\bigr)^{\ast}d_1^{e_mg_j}\rho_{e_ke_m} \nonumber \\ 
&- \underset{e_k,g_m}{\sum }\Bigl[\Gamma_{p,e_kg_j}^{\ast} \bigl(d_1^{g_ie_k}\bigr)^{\ast}d_1^{e_kg_m}\rho_{g_mg_j} \nonumber \\
&+\Gamma_{p,g_ie_k} \bigl(d_1^{g_me_k}\bigr)^{\ast}d_1^{e_kg_j}\rho_{g_ig_m}\Bigr] - i\omega_{g_ig_j}\rho_{g_ig_j} \nonumber \\
&+ \underset{e_i, e_j}{\sum}\Gamma_{g_ig_j}^{e_ie_j}\rho_{e_ie_j} -\gamma\rho_{g_ig_j} + \lambda\delta\bigl(g_i, g_j\bigr)
\label{rate1}
\end{align}
and
\begin{align}
\dfrac{\partial\rho_{e_ie_j}}{\partial t} =&\bigl(\Gamma_{p,e_ig_m}^{\ast} + \Gamma_{p,g_ke_j}\bigr) \underset{g_k, g_
m}{\sum }d_1^{e_ig_k}\bigl(d_1^{g_me_j}\bigr)^{\ast}\rho_{g_kg_m} \nonumber \\ 
&- \underset{g_k,e_m}{\sum }\Bigl[\Gamma_{p,g_ke_j} d_1^{e_ig_k}\bigl(d_1^{g_ke_m}\bigr)^{\ast}\rho_{e_me_j} \nonumber \\
&+\Gamma_{p,e_ig_k}^{\ast} d_1^{e_mg_k}\bigl(d_1^{g_ke_j}\bigr)^{\ast}\rho_{e_ie_m}\Bigr]  \nonumber \\
&- i\omega_{e_ie_j}\rho_{e_ie_j} - \Gamma\rho_{e_ie_j}
\label{rate2},
\end{align}
where $\rho_{g_ig_j}$ and $\rho_{e_ie_j}$ are the density matrix elements for the ground and excited states, respectively. The first term in (\ref{rate1}) describes the re-population of the ground state and the creation of Zeeman coherences due to induced transitions, $\Gamma_{p,g_ie_j}$ and $\Gamma_{p,e_ig_j}^{\ast}$ represent the laser field coupling of the ground and excited states, and $d_1^{e_ig_j}$ is the dipole transition matrix element. The second term stands for the changes of ground state Zeeman sublevel population and the creation of ground state Zeeman coherences due to light absorption. The third term describes the destruction of the ground state Zeeman coherences by the external magnetic field; $\omega_{g_ig_j}$ is the splitting of the ground state Zeeman sublevels. The fourth term describes the re-population and transfer of excited state coherences to the ground state due to spontaneous transitions. We assume our transition to be closed (within the hyperfine structure), and so $\underset{e_i, e_j}{\sum}\Gamma_{g_ig_j}^{e_ie_j} = \Gamma$. The fifth and sixth terms show the relaxation and re-population of the ground state due to non-optical reasons - in our case it is assumed to be solely transit relaxation. It is assumed that the atomic equilibrium density outside the interaction region is normalized to 1, therefore $\lambda = \gamma$.

In  equation (\ref{rate2}) the first term stands for the light absorbing transitions; the second term denotes induced transitions to the ground state; the third describes the destruction of ground state Zeeman coherences in the external magnetic field; and the fourth term denotes the rate of spontaneous decay of the excited state. where $\omega_{e_ie_j}$ is the splitting of the excited state Zeeman sublevels.

The interaction strength $\Gamma_{p,g_ie_j}$ is given by
\begin{equation}
 \Gamma_{p,g_ie_j} = \frac{|\varepsilon_{\overline{\omega}}|^2}{\hbar^2}\frac{1}{\left[\left(\frac{\Gamma}{2} + \frac{\Delta\omega}{2}\right) \pm i\left(\overline{\omega} - \boldsymbol{k}_{\overline{\omega}}\boldsymbol{v} - \omega_{e_jg_i}\right)\right]}
 \label{gammaP1},
\end{equation}
where $\frac{|\varepsilon_{\overline{\omega}}|^2}{\hbar^2}$ is proportional to the laser power density.

The magnetic field not only leads to Zeeman sublevel splitting $\omega_{ij}$, but changes the transition dipole elements due to magnetic sublevel mixing. In the case of two hyperfine structure levels this mixing can be obtained from the Breit-Rabi formula~\cite{Breit:1931, Aleksandrov:1993}.

We are conducting experiments in stationary excitation conditions so that $\partial\rho_{g_ig_j}/\partial t = \partial\rho_{e_ie_j}/\partial t = 0$. Thus, the differential equations (\ref{rate1}) + (\ref{rate2}) are reduced to a system of linear equations, which, when solved, yields the density matrices for the atomic ground and excited states. The observed fluorescence intensity can be calculated as follows:
\begin{equation}
        I_f(\tilde{\textbf{e}})=\tilde{I}_0\underset{g_i, e_i, e_j}{\sum}d^{(ob)*}_{g_ie_j}d^{(ob)}_{g_ie_i}\rho_{e_ie_j},
        \label{fluorescence}
\end{equation}
where $\tilde{I}_0$ is a constant of proportionality.

To include all the different velocity groups of atoms, we average over the Doppler profile simply by calculating the intensity for each velocity group and multiplying this intensity by the respective statistical weight corresponding to the number of atoms in the respective velocity group. Also, to obtain the total signal for a specific observation direction, we average the fluorescence signal over the two orthogonal polarization components. Finally, since we do not resolve the fluorescence spectrally, we sum the signal over all four possible radiating hyperfine transitions.

There are two parameters in the model which need to be determined by comparing the results from calculations with experiment: (1) the conversion between laser power density $I$ and Rabi frequency $\Omega$ and (2) the conversion between beam diameter and transit relaxation time $\gamma^{-1}$. Both of them can be theoretically estimated~\cite{Alnis:2003,Pfleghaar:1993}, though the estimate of the conversion factor in the case of the Rabi frequency is rather rough and needed to be adjusted by a factor of less than two to yield the best match between theory and experiment. The transit relaxation rate can be estimated from the mean thermal velocity of the atoms $v_{avg}$ and the laser beam diameter $d$ (FWHM), and is close to $v_{avg}/d$. Since we assumed that $\Omega \propto \sqrt{I}$ and $\gamma \propto 1/d$, the same values for these two conversion factors were used in all our measurements, which, for two rubidium isotopes, all together spanned 8 hyperfine transition and a large range of laser power densities and beam diameters. We also used 293 K for the temperature, 10 MHz for the laser linewidth. Atomic constants such as nuclear spin Land\'e factor and the hyperfine structure constant $A$ were taken from~\cite{steck:rubidium85,steck:rubidium87}. Hyperfine structure constants $B$ were assumed to be negligibly small.

\section{\label{Results:level1}Results and Discussion}
\begin{figure*}[htbp]
	\centering
		\resizebox{\textwidth}{!}{\includegraphics{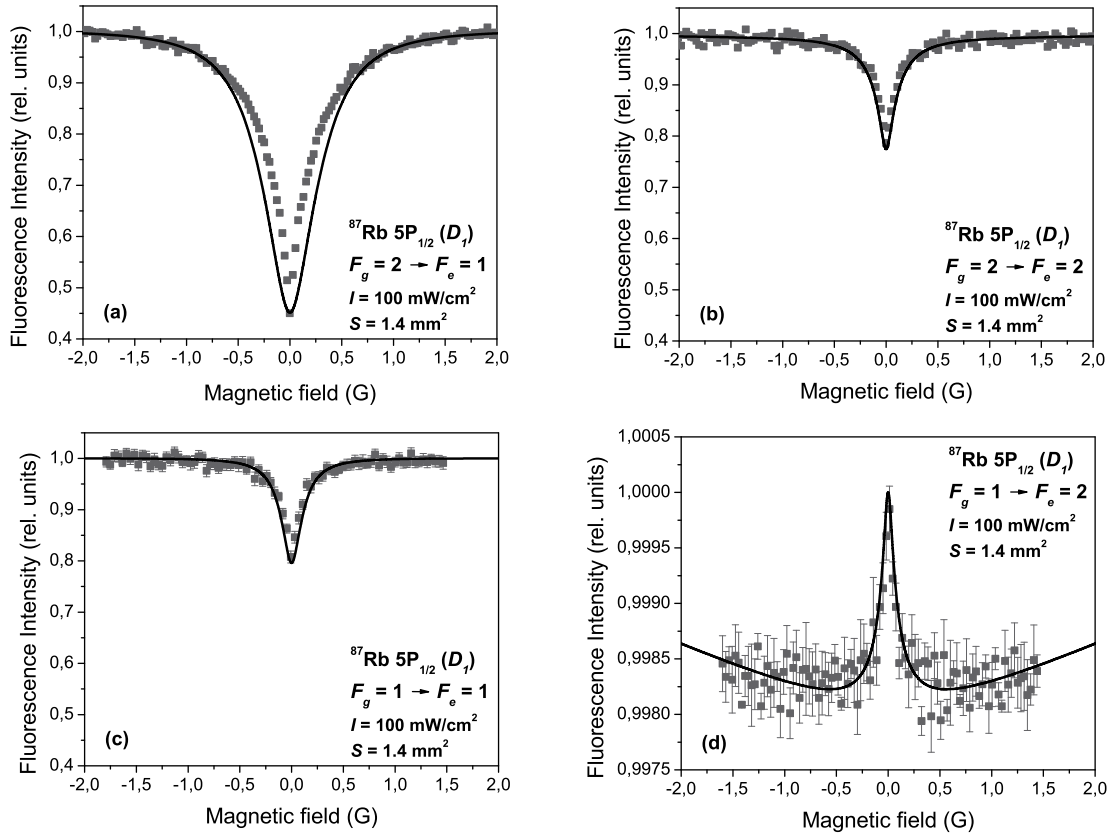}}
	\caption{\label{fig:resonances87} Fluorescence intensity versus magnetic field for $^{87}$Rb at $D_1$ excitation. Filled squares, experiment; solid line, theory. Note that the vertical axis is identical for (a)--(c), but markedly differs in (d). The excited state, total angular momentum of the ground $F_g$ and excited states $F_e$ of the transition, laser power density $I$, and laser beam cross-section (FWHM) $S$ are given in each panel. 
}
\end{figure*}

\begin{figure*}[htbp]
	\centering
		\resizebox{\textwidth}{!}{\includegraphics{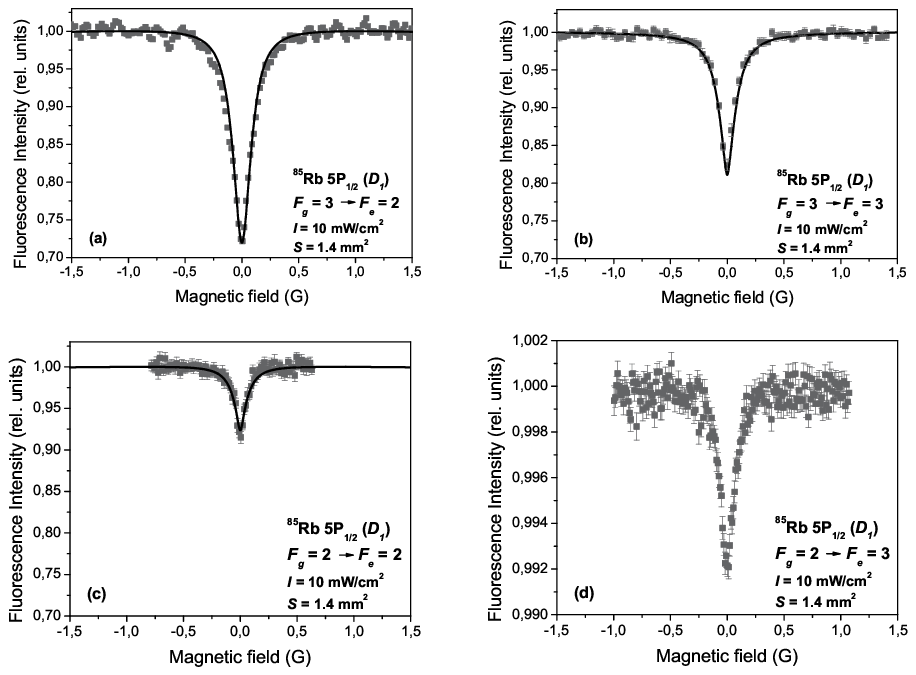}}
	\caption{\label{fig:resonances85} Fluorescence intensity versus magnetic field for $^{85}$Rb at $D_1$ excitation. Filled squares, experiment; solid line, theory. Note that the vertical axis is identical for (a)--(c), but markedly differs in (d). The excited state, total angular momentum of the ground $F_g$ and excited states $F_e$ of the transition, laser power density $I$, and laser beam cross-section (FWHM) $S$ are given in each panel. 
}
\end{figure*}

Figures~\ref{fig:resonances87} and~\ref{fig:resonances85} show magneto-optical resonances at laser excitation frequencies tuned to the corresponding hyperfine transitions of the $D_1$ line of rubidium. Figure~\ref{fig:resonances87} shows the transitions of $^{87}$Rb at 100 mW/cm$^2$, whereas Fig.~\ref{fig:resonances85} shows the transitions of $^{85}$Rb at 10 mW/cm$^{2}$. Experimental measurements are represented by markers, whereas the results of theoretical calculations are shown by solid lines. In each figure, panels (a)--(c) have the same vertical scale, but the vertical scale of panel (d) is much smaller. In most cases, the theoretical model reproduces the measured signals to within experimental accuracy. Only at high laser power densities ($\geq$ 100 mW/cm$^2$), some discrepancies arise as regards the resonance widths, although the model predicts the resonance contrast quite accurately even at high laser power densities. Indeed, in Figure~\ref{fig:resonances87}(a), which describes the $^{87}$Rb $F_g=2 \rightarrow F_e=1$ transition, the calculations predict a somewhat larger resonance width than is observed. This transition is the strongest transition at a rather high laser power density, and so absorption saturation effects could be strong enough at the wings of the spatial beam profile. At strong saturation of the absorption, it is no longer adequate to describe the transit relaxation rate by a single rate constant~\cite{Auzinsh:1983,Auzinsh:2008}, as is done in our theoretical model, since the resonance width is determined mainly by the ground state relaxation rate, which is dominated by the transit relaxation rate in this experiment.

\begin{figure*}[htbp]
	\centering
		\resizebox{\textwidth}{!}{\includegraphics{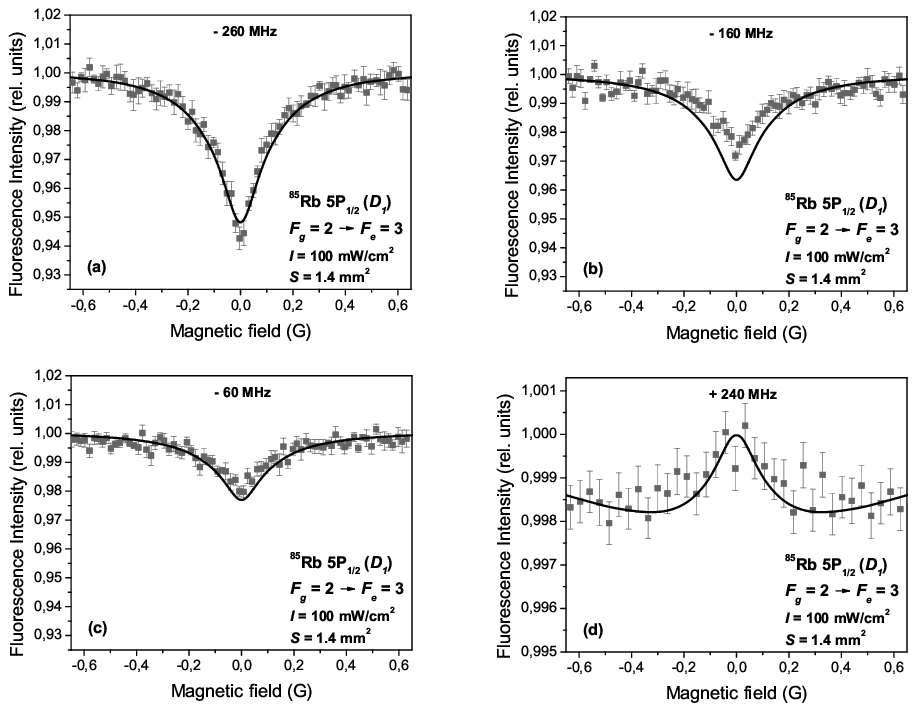}}
	\caption{\label{fig:detuning} Fluorescence intensity versus magnetic field for the $F_g=2\rightarrow F_e=3$ transition of $^{85}$Rb at various values of the laser detuning with respect to the exact position of the $F_g=2\rightarrow F_e=3$ transition. Filled squares, experiment; solid line, theory.}
\end{figure*}

	Figures~\ref{fig:resonances87}(a)--(c) and~\ref{fig:resonances85}(a)--(c) show dark resonances, which are expected, since in each case ($F_g \geq F_e$). The contrasts for the two cases where $F_g=F_e$ (Fig.~\ref{fig:resonances87}(b) and (c)) are equal, whereas when $F_g>F_e$ (Fig.~\ref{fig:resonances87}(a)) the contrast is larger by more than a factor of two. Similar behavior is seen in Figure~\ref{fig:resonances85}, although in this case the two transitions with $F_g=F_e$ do not have similar contrasts, because adjacent resonances are not resolved. The $F_g=3\rightarrow F_e=3$ transition (Fig.~\ref{fig:resonances85}(b)) is a dark resonance that is adjacent to another dark resonance at $F_g=3\rightarrow F_e=2$ (Fig.~\ref{fig:resonances85}(a)), and hence its contrast is greater than for the resonance at the $F_g=2\rightarrow F_e=2$ transition (Fig.~\ref{fig:resonances85}(c)), which is a dark resonance that is next to a resonance at $F_g=2\rightarrow F_e=3$, which is expected to be bright. Figure~\ref{fig:resonances87}(d) shows a bright resonance, which is also expected because  $F_g<F_e$. The contrast of the resonance in Fig.~\ref{fig:resonances87}(d) is about 0.15\%, but the agreement between experiment and calculation is still quite good.  Figure~\ref{fig:resonances85}(d) describes the  $^{87}$Rb $F_g=2 \rightarrow F_e=3$ transition, which should also be a bright resonance because $F_g<F_e$. However, the measured curve, which was recorded at the laser frequency that corresponded to maximum LIF, shows a dark resonance. We can understand why by studying Figure~\ref{fig:detuning}. 

	Figure~\ref{fig:detuning}(a)--(d) shows resonances obtained when the laser detuning with respect to the $F_g=2\rightarrow F_e=3$ transition was $-260$ MHz, $-160$ MHz,  $-60$ MHz, and $+240$ MHz, respectively.  A detuning of $-360$ MHz with respect to the $F_g=2\rightarrow F_e=3$ transition would correspond exactly to the $F_g=2\rightarrow F_e=2$ transition, which is expected to be dark and is shown in Figure~\ref{fig:resonances85}(c). The dark resonance at $F_g=2\rightarrow F_e=2$ transition is clearly much stronger than the bright resonance at the $F_g=2\rightarrow F_e=3$ transition, and overwhelms the bright resonance when the laser is tuned exactly to the  $F_g=2\rightarrow F_e=3$ transition, or to the maximum in LIF near that transition at a slightly lower frequency. However, as can be seen in Fig.~\ref{fig:detuning}, when the laser is tuned past the  $F_g=2\rightarrow F_e=3$ transition, further away from the  $F_g=2\rightarrow F_e=2$ transition, a bright resonance can be observed. The respective theoretical calculations show the same behavior with good agreement. As far as we know such a detailed comparison of calculated and measured non-linear magneto-optical resonance profiles as a function of laser detuning has not been presented previously, and it is satisfying to note that a careful theoretical treatment can reproduce well the resonance profiles at an unresolved transition for laser excitation frequencies that differ by only 100 MHz.
  
\begin{figure*}[htbp]
	\centering
		\resizebox{\textwidth}{!}{\includegraphics{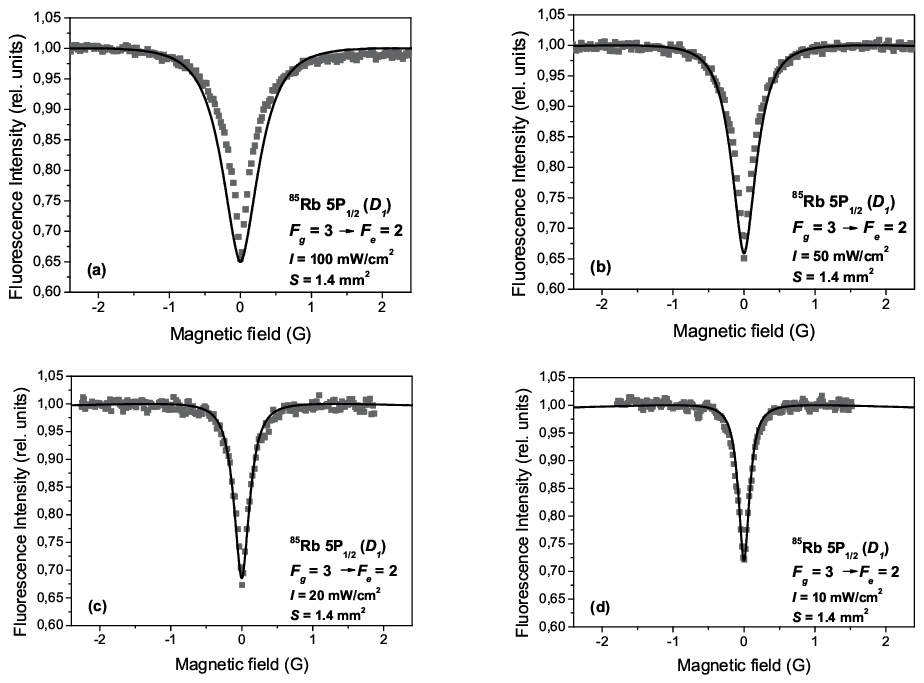}}
	\caption{\label{fig:dark} Resonance signals for $^{85}$Rb at the $F_g=3\rightarrow F_e=2$ transition for different laser power densities $I$. Filled squares, experiment; solid line, theory.}
\end{figure*}

\begin{figure*}[htbp]
	\centering
		\resizebox{8cm}{6cm}{\includegraphics{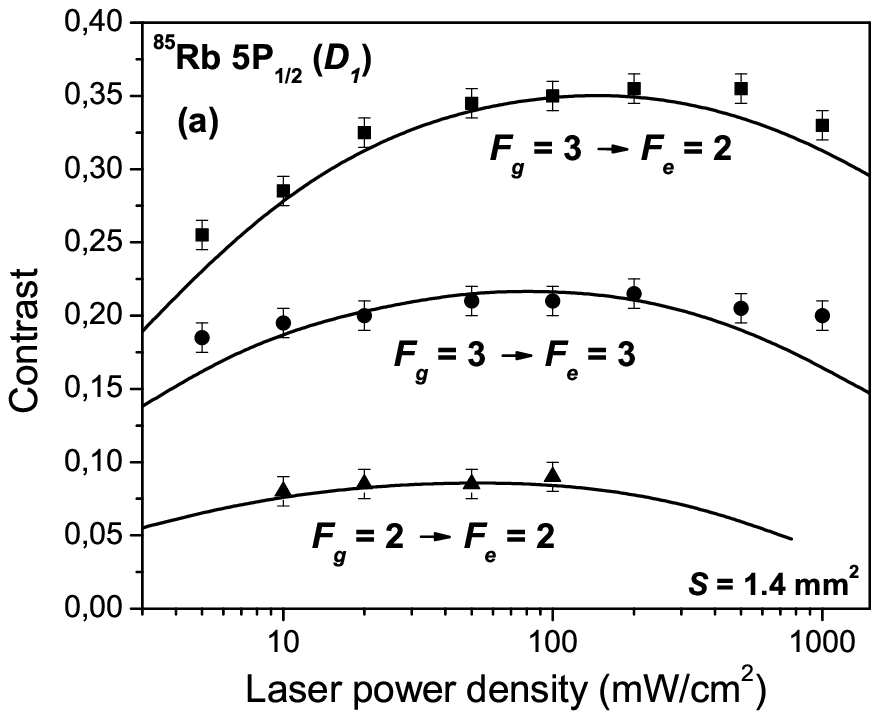}}
		\resizebox{8cm}{6cm}{\includegraphics{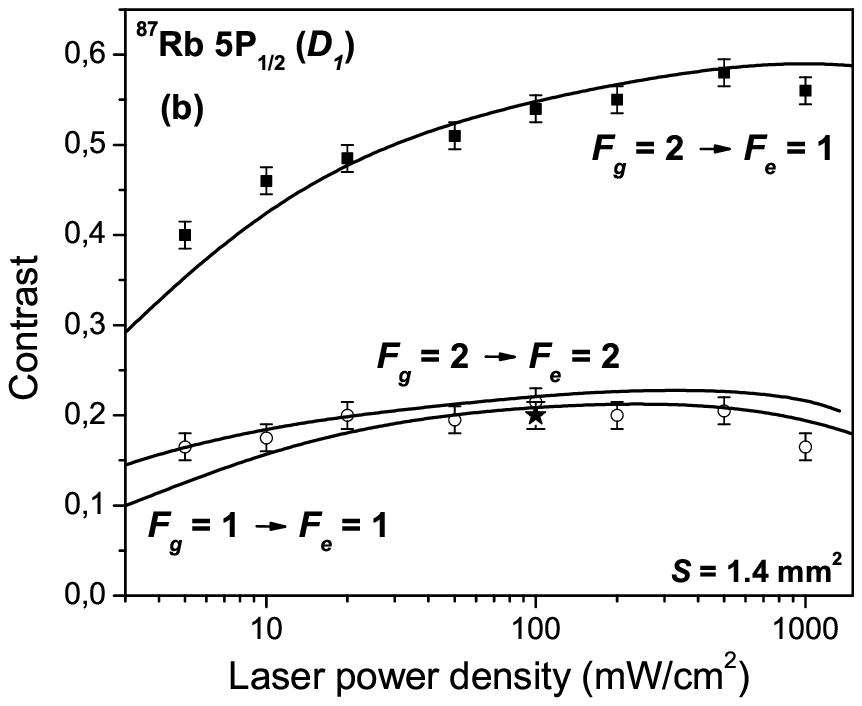}}
		\resizebox{8cm}{6cm}{\includegraphics{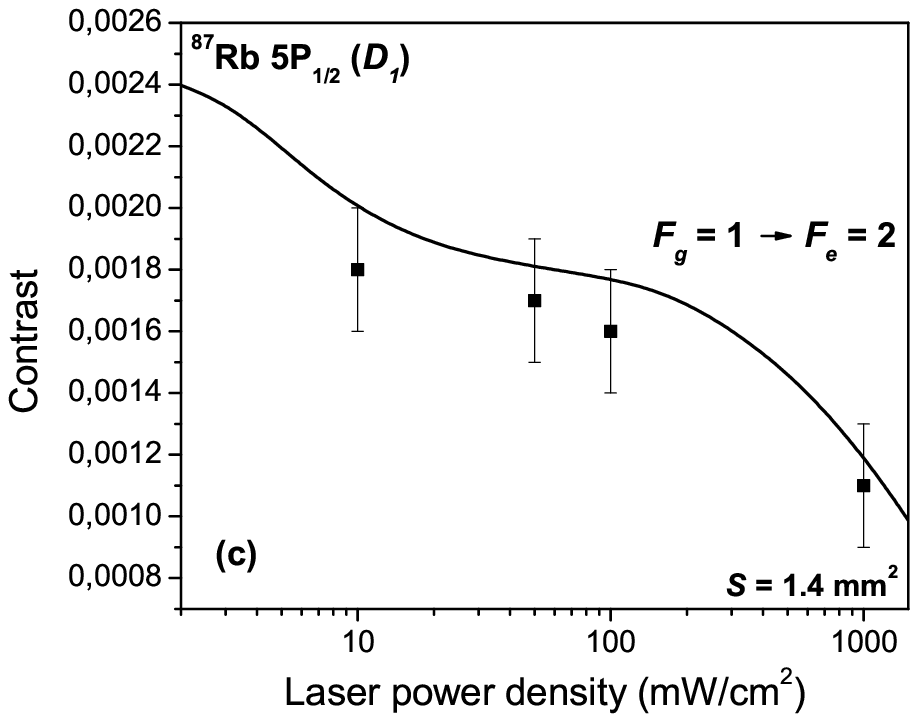}}
	\caption{\label{fig:contrast} Resonance contrast as a function of laser power density for  $^{85}$Rb dark resonances (a), $^{87}$Rb dark resonances (b), and $^{87}$Rb bright resonance (c); beam cross section was $S=1.4$ mm$^2$. Markers, experiment; solid line, theory. The solid star in (b) is the single measurement point for the $F_g=1\rightarrow F_e=1$ transition. The open circles represent the results from measurements for the $F_g=2\rightarrow F_e=2$ transition.}
\end{figure*}

	We also studied how the resonance shape varies as a function of laser power density. Figure~\ref{fig:dark} shows resonance signals for $^{85}$Rb when the $F_g=3\rightarrow F_e=2$ transition is excited at different laser power densities. The dependence of contrast on laser power density for all transitions can be extracted from plots similar to the ones in Fig.~\ref{fig:dark}, and presented as in Fig.~\ref{fig:contrast}, which shows the resonance contrast as a function of laser power density for the dark resonances of (a)$^{85}$Rb and (b)$^{87}$Rb and (c) for the bright resonance at the $F_g=1\rightarrow F_e=2$ transition of $^{87}$Rb. The contrast $C$ is defined as $C=\lvert I_{res}-I_{far}\rvert/I_{far}$, where $I_{res}$ is the LIF intensity at the resonance at zero magnetic field and $I_{far}$ is the LIF intensity far from the resonance, where the curve of LIF versus magnetic field becomes flat or has an extremum. Markers represent experimentally measured points, and lines represent the results of theoretical calculations. It can be seen that the model predicts the resonance contrast very well over more than two orders of magnitude in laser power density, and even at laser power densities at which one of the model's assumptions is not expected to be fully adequate, namely, the validity of describing the transit relaxation by a single rate constant~\cite{Auzinsh:1983}. 

\begin{figure*}[htbp]
	\centering
%		\resizebox{\textwidth}{!}{\includegraphics{Fig6}}
		\resizebox{8cm}{6cm}{\includegraphics{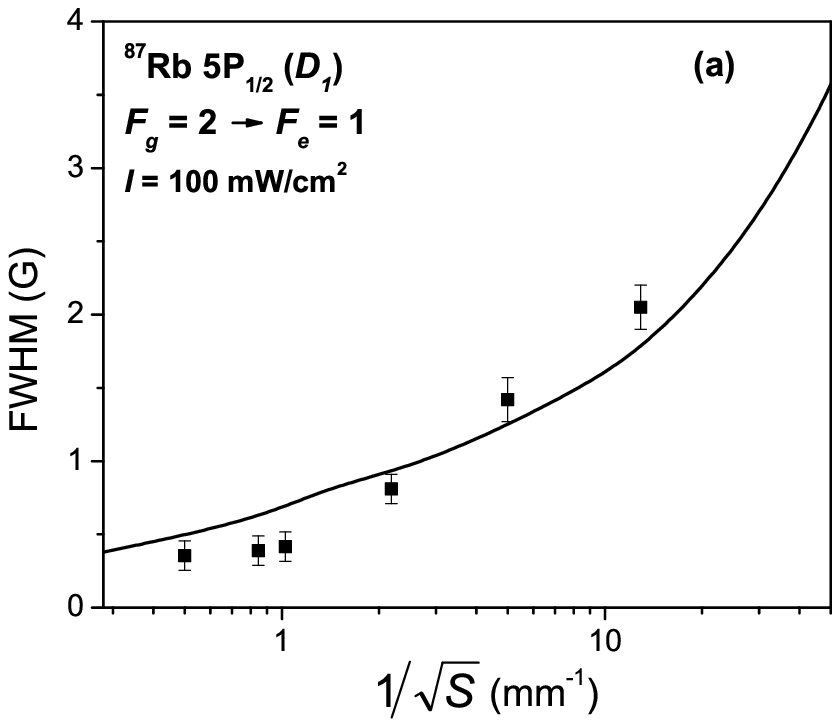}}
		\resizebox{8cm}{6cm}{\includegraphics{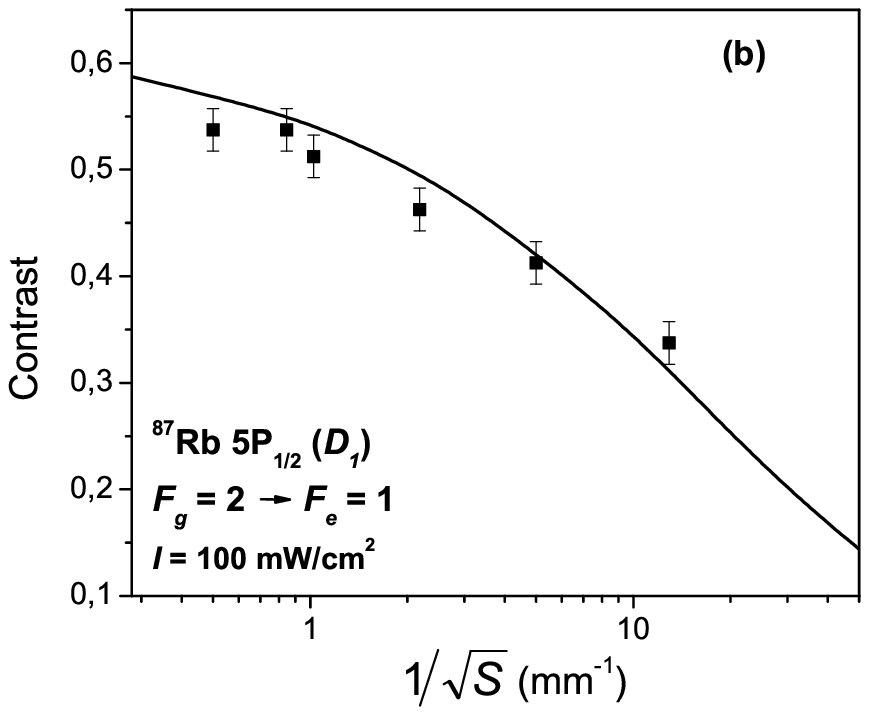}}
%		\resizebox{8cm}{6cm}{\includegraphics{Fig6b}}
	\caption{\label{fig:beam} Resonance width versus the inverse of the square root of the laser beam cross-sectional area $S$ (FWHM) for the $F_g=2\rightarrow F_e=1$ transition of $^{87}$Rb. Filled squares, experiment; solid line, theory.}
\end{figure*}

	Figure~\ref{fig:beam} shows the results of a study of how the resonance contrast and full width at half maximum (FWHM) depend on the transit relaxation rate. Transit relaxation is the process whereby atoms relax by leaving the laser beam. Thus, the transit relaxation rate is proportional to the inverse of the beam diameter. In Fig.~\ref{fig:beam}, we plot the resonance contrast and full width at half maximum (FWHM) as a function of the inverse square root of the cross-sectional area of the beam, which is essentially the inverse of the beam diameter. The results are for the $F_g=2\rightarrow F_e=1$ transition of $^{87}$Rb. Again markers show the results of experimental measurements, whereas the line shows the result of calculations. The resonance width increases monotonically with relaxation rate, and the calculations predict the resonance width reasonably well over the range of relaxation rates studied. 

%\section{\label{Analysis:level1}Analysis and Discussion}

\section{\label{Conclusion:level1}Conclusion}
The bright and dark magneto-optical resonances $D_1$ transition of natural rubidium have been carefully measured with special efforts to eliminate systematic errors, and a detailed theoretical treatment has been applied to describe the experimental results in a system that is only partially resolved under Doppler broadening. The model was based on the optical Bloch equations for the density matrix and included averaging over the Doppler profile, mixing of magnetic sublevels in an external magnetic field, participation of all nearby transitions, and a detailed treatment of the coherence properties of the exciting laser radiation. The calculation based on our model accurately reproduces the experimental signals at each hyperfine transition and over a wide range of laser power densities and beam diameters. A bright resonance was observed at the  $F_g=2\rightarrow F_e=3$ transition of $^{85}$Rb, but displaced from the exact position of the transition due to the influence of the nearby  $F_g=2\rightarrow F_e=2$ transition, which is a dark resonance whose contrast is almost two orders of magnitude larger than the contrast of the bright resonance at the $F_g=2\rightarrow F_e=3$ transition. The amplitude of the bright resonance is so small because the bright resonance is formed by a bright state that is strongly and continuously coupled to the laser field in a partially open transition, and as a result, this population of atoms in this state is heavily depleted. In contrast, the dark resonance is formed by a dark state which is completely decoupled from the light field and which can efficiently maintain its population even for an open transition. Even in this very delicate situation, the theoretical model described the experimental signals quite well at different laser detunings. As far as we know, the the  bright resonance expected at the $F_g=1\rightarrow F_e=2$ transition of $^{87}$Rb had not been observed before. 

We believe that this study tests the theoretical model based on the optical Bloch equations to its fullest potential because experiment and theory were compared for all eight hyperfine transitions of two isotopes at various laser power densities, laser beam diameters and laser frequencies. Agreement between experiment and theory over this wide range of situations and experimental conditions was excellent. Previous studies that presented comparisons between experiment and theory for magneto-optical resonances in similar atomic systems have not had to take into account the Doppler effect because they used atomic beams~\cite{Picque:1978, Renzoni:1997}, worked in systems resolved under Doppler broadening~\cite{Auzinsh:2008}, have not presented comparisons of theory and experiment over such a wide range of situations~\cite{Andreeva:2002}, or have not included in the theoretical model a detailed treatment of the coherence properties of the laser radiation or effects such as mixing of magnetic sublevels in an external magnetic field~\cite{Andreeva:2002}.

\begin{acknowledgments}
We would like to thank Maris Tamanis for invaluable assistance with the experiments. We thank Robert Kalendarev for preparing one of the rubidium cells and Christina Andreeva and Antoine Weis for useful discussions. We are grateful for support from the Latvian National Research Programme in Material Sciences Grant No. 1-23/50, the University of Latvia Grant No. Y2-ZP04-100, and the Latvian Science Council Grant No. LZP 09.1196. A.~J., F.~G., and L.~K. gratefully acknowledge support from the ESF project. 
\end{acknowledgments}
\bibliography{rubidium}% Produces the bibliography via BibTeX.
\end{document}